\documentstyle[aps,epsf,multicol,psfig]{revtex}

\begin{document}

\draft

\sloppy

\title{Temperature dependence of the microwave surface impedance
measured on different kinds of MgB$_2$ samples.}
\author{A.A.~Zhukov\dag, L.F.~Cohen, A. Purnell, Y.~Bugoslavsky \ddag,
A.~Berenov, and J.L.~MacManus-Driscoll}
\address{Center for High Temperature Superconductivity,
Imperial College of Science Technology and Medicine, Prince
Consort Road, London SW7 2BZ, U.K.}
\address{\dag Institute of Solid State Physics, Russian Academy
of Science, Chernogolovka, 142432 Russia}
\address{\ddag General Physics Institute, Moscow, Russia}
\author{H.Y~Zhai, Hans M. Christen, Mariappan P. Paranthaman and Douglas H.
Lowndes}
\address{Oak Ridge National Laboratory, Oak Ridge, TN 37931-6056}
\author{M.A. Jo and M.C. Blamire}
\address{Department of Materials Science, Cambridge University,
Cambridge CB2 3QZ}
\author{Ling~Hao and J.~Gallop}
\address{National Physical Laboratory, Queens Rd, Teddington, UK}
\date{July 11, 2001}
\maketitle

\begin{abstract}
In this paper we present the results of measurements of the
microwave surface impedance of a powder sample and two films of
MgB$_2$. One film has $T_c=30$~K and is not textured, the other is
partially c-axis orientated with $T_c=38$~K. These samples show
different types of temperature dependence of the field penetration
depth: linear for the powder sample, exponential with
$\Delta/kT_c<1.76$ (film with $T_c=30$~K) and strong coupling
behavior with $\Delta/kT_c\sim 2.25$ (film with $T_c=38$~K). The
results are well described in terms of an anisotropic gap model or
by the presence of an Mg deficient phase.
\end{abstract}
\vskip 0.05in
\begin{multicols}{2}
\newpage
Since the discovery of superconductivity in MgB$_2$  \cite{Naga} a
number of papers have been dedicated to the measurement of the
energy gap. Determining whether MgB$_2$ resembles conventional
superconductors is of primary importance. A variety of techniques
have been employed such as tunnelling, Raman, specific heat, etc.
(see. Table I). The first tunnelling measurements were performed
on bulk polycrystalline material with values of the gap varying
from 2 meV \cite{Rubio} to 8 meV \cite{Bugoslavsky}. In the latest
tunneling \cite{Giubileo2} and point contacts \cite{Szabo}
experiments coexistence of two-gap structure ($\Delta_L(T)$ and
$\Delta_S(T)$) was observed up to the transition temperature. A
two-gap structure was also observed in photoemission \cite{Tsuda}.
Specific heat measurements of bulk polycrystalline material
\cite{Junod,Bouquet} show that it is necessary to involve either
two gaps or a single anisotropic gap $\Delta(\vec k)$ \cite{Haas}
($\Delta_S=min(\Delta(\vec k))$, $\Delta_L=max(\Delta(\vec k))$,
where $\vec k$ belongs to the Fermi surface) to describe the data.
The sign of the anisotropic gap was obtained from tunnelling
\cite{Chen} into a c-axis oriented film \cite{Moon}. The size of
the smaller gap $\Delta_{ab}\simeq 5$ meV was supported by far
infrared conductivity (FIR) \cite{Jung} where a single gap
$\Delta=5.2$ meV was observed on the same type of film. There are
no reported observations of the two gap structure in tunnelling or
point contact measurements made in films, as far as we know. At
this point the key question is whether the material has an
isotropic, an anisotropic gap or multiple gap structure.

The value of the superconducting gap can be determined by
measuring the temperature dependence of the field penetration
depth ($\lambda(T)$). $\lambda(T)$ is connected to the
concentration of the superconducting carriers and thus to the
structure of the superconducting gap. Most of the reported data
can be fitted with an analytical formula
$\lambda(T)=\lambda_0/(1-(T/T_c)^n)^{1/2}$ with different values
of $n$ \cite{Chen,Li,Panagopoulos,Pronin}. For the reported values
of $n$ and details of the experiments see Table II, all
experiments were performed on polycrystalline samples. These
temperature behaviors do not contradict the two-gap model with
s-wave order parameter, and values of $n=1\div 2$ can be
interpreted as the presence of the small gap
($\Delta_S/kT_c<1.76)$ \cite{Golubov}. The value of $\Delta_S=2.8$
meV was extracted from AC susceptibility of especially prepared
powder samples \cite{Manzano}.

Measurements of the temperature dependence of the real part of the
microwave surface impedance $Z_s(T)=R_s(T)+i\omega\mu_0\lambda(T)$
\cite{Zhukov,Hakim} demonstrated that $\Delta R_s(T)\propto T$ at
low temperatures. This behavior is associated with the presence of
the small gap also.

In this paper we report measurements of $Z_s(T)$ performed on
three types of MgB$_2$ sample: a powder sample, a randomly
oriented polycrystalline film with $T_c=30$K (Film~I) and c-axis
orientated film with $T_c=38$K (Film~II). We would like to stress
that as well as determining $\lambda(T)$, the microwave
measurement checks the quality of the sample under investigation
by using as a parameter the value of residual surface resistance
$R_{res}=R_s(T=0)$. Linear temperature behavior of the deviation
of the field penetration depth $\Delta\lambda(T)$ on the powder
sample was observed. $\Delta\lambda(T)$ measured on the films
demonstrates activated behavior with $\Delta/kT_c<1.76$ for the
Film~I and $\Delta/kT_c\simeq 2.25$ for the Film~II.

In our measurements we used a multilayered powder (Alfa Aesar Co.,
98\% purity) sample fixed in wax (as described in ref.
\cite{Zhukov} sample 3). Film I (4mm$\times$ 6mm$\times$ 300nm)
was prepared by pulsed laser deposition (PLD) on c-plane oriented
Al$_2$O$_3$ substrate with post annealing in ArH$_2$ atmosphere
and has $T_c=30$~K (see \cite{Berenov}, film CAM 6). Film II
partially c-axis orientated (4.5mm$\times$ 10mm$\times$ 750nm) was
prepared by e-beam evaporation on R-plane Al$_2$O$_3$ and then
annealed in Mg vapor for 1 hour (see \cite{Zhai} and reference
therein, film A type). DC resistivity measurements result
$\rho(300$~K$)=250 \mu\Omega\cdot$cm and $\rho(T_c)=250
\mu\Omega\cdot$cm for the Film I. The width of the superconducting
transition of this film is $\delta T_c=5$~K. For the Film II these
values are $\rho(300$~K$)=16\mu\Omega\cdot$cm,
$\rho(T_c)=6\mu\Omega\cdot$cm and $\delta T_c=0.3$~K
\cite{Paranthaman}. For the powder sample
$\rho(T_c)=25\mu\Omega\cdot$cm was calculated from the surface
resistance measurement. A crude evaluation can be made of the mean
free path for Film II using data from the Hall measurements
($R_H(T_c)=5.5\cdot 10^{-11}$m$^3$/C) performed on the same type
of film \cite{Jin} and from the calculated mean value of the Fermi
velocity $v_F=4.8\cdot 10^5$m/s \cite{Kortus}, such that
$l(T_c$,Film II$)\simeq3($nm$)\sim\xi$. Other samples are in dirty
limit because of the higher values of the resistivity.

Experiments were performed in two similar dielectric puck
resonator and copper housing systems operated at different
frequencies. Measurements of $R_s(T)$ of the two films are made
using a TiO$_2$ dielectric puck resonator at 3.5~GHz. Measurements
of the $\Delta\lambda(T)$ of all samples are made using an alumina
dielectric puck resonator at 8.8~GHz to minimize errors in
temperature dependence of the frequency shift. The fundamental
TE$_{011}$ mode was used for all measurements, details of
experimental set up are given elsewhere \cite{Karen}. Film and
powder samples were attached directly to the puck or to the quartz
spacer with Apiezon N grease. Absolute values of $Z_s$ were
extracted using the relation
$R_s(T)+i\omega\mu_0\Delta\lambda(T)=G_s(1/Q(T)-
1/Q_0(T)+i(f_0(T)- f(T))/2f)$, where $1/Q_0(T)$, $1/Q(T)$,
$f_0(T)$ and $f(T)$ are quality factors and frequency shifts of
unloaded resonator and resonator with the sample. Calculations of
the geometry factor ($G_s$) were performed with MAFIA (commercial
available software) \cite{Hao}.


Fig. \ref{Rs} presents $R_s(T)$ data at low temperatures of all
three samples (powder (solid circles), Film I (open circles) and
Film II (solid squares)) rescaled to the 3.5~GHz assuming an
$\omega^2$ law (measurements of the $R_s(T)$ of these films at
different frequencies using dielectric puck and parallel-plate
resonators will be published elsewhere \cite{Zhukov2}). Two
distinct behaviors of $R_s(T)$ are seen on the graph. Powder and
Film I demonstrate quite high residual loses
$R_{res}=R_s(T=10$K$)=550\mu\Omega$ and 570$\mu\Omega$, and
subsequently show much steeper slopes in comparison with Film II
($R_{res}=110\mu\Omega$). Such a low residual surface resistance
of the Film II is comparable to the values obtained in the MgB$_2$
wire $R_s(20$K$)=57\mu\Omega$ \cite{Hakim} and an ion milled film
$R_s(20$K$)=92\mu\Omega$ \cite{Lee}. Values are converted to
3.5~GHz assuming an $\omega^2$ law. As was mentioned in our
previous paper, the steep slope of $R_s(T)$ can be attributed to
the presence of low energy exitations caused by the small
superconducting gap.

Fig. \ref{Xs1} shows the temperature dependence of the deviation
of the field penetration depth ($|Delta\lambda(T)$) of the powder
sample (open circles) and Film I (solid circles). Linear
temperature dependence of $\Delta\lambda(T)$ of powder sample from
$T=10$~K is observed. Temperature dependence of the
$\Delta\lambda$ of the Film I (solid line) can be fitted well by
the formula

\begin{eqnarray}
\label{Xs1weak} \Delta\lambda(T)={\lambda _0\over\left[ {1-\left(
{T\over T_c} \right)^{n}} \right]^{0.5}}\coth \left[ {{d \over
{\lambda _0\over\left[ {1-\left( {T\over T_c} \right)^{n}}
\right]^{0.5}}}} \right]\nonumber\\
-\lambda _0\coth \left[ {{d \over {\lambda _0}}} \right]
\end{eqnarray}
where $d=300$ nm is the thickness of the film and $n=3-T/T_c$
\cite{Bonn} describes the weak coupling regime in the clean limit.
The extracted value $\lambda_0=300\pm 20$~nm is associated with
changes in screening originating only from Copper pairs belonging
to the part of Fermi surface with the "smaller gap" and cannot be
associated with London penetration depth $\lambda_L(0)$. The value
of the gap obtained from the weak coupling fit is an
overestimation and it is definitely less than the BCS value
$1.76kT_c$ because Film I is in dirty limit. To extract the exact
value of the smaller gap observed on Film I additional
calculations are necessary. The form of $\Delta\lambda(T)$ in the
strong coupling limit ($n=4$) with the same values of $d$ and
$\lambda_0$ demonstrates disagreement with the data in the whole
range of temperatures (dashed line).

Fig. \ref{Xs2} shows $\Delta\lambda(T)$ measured on Film II
(dots). The solid line shown on the graph is the fit to the
equation (1) with $n=4$ and $d=750$ nm. From the fit the extracted
value of $\lambda_L(0)$ is $110\pm 10$~nm. This value is less than
most reported (140~nm \cite{Finnemore}, 180~nm \cite{Chen2}, 132.5
nm \cite{Li}, 110 nm \cite{Thompson}, 85 nm \cite{Panagopoulos},
160 nm \cite{Manzano}). The extracted value of $\Delta(0)$ is
$7.4\pm 0.25$ meV, where the dash-dotted line is the fit
$\Delta\lambda(T)=\lambda_0(2.25\pi T_c/2T)^{0.5}exp(-
2.25T/T_c)$, which is in good agreement with our data up to
22.5~K$\simeq T_c/2$. Fitting curve of the weak coupling regime
(dashed line) shows clear disagreement with our data. We do not
think that the smaller gap could be suppressed or masked in Film
II. In addition the low $R_{res}$ means that we can exclude the
coexistence of two separate gaps in this film, with some
confidence.

Thus in three different kind of samples we observed two different
kind of $\Delta\lambda(T)$. Bulk material with $T_c=39$~K shows
linear temperature dependence associated with the presence of the
small gap. This data is in accordance with measurements performed
on polycrystalline material (see Table I). $\Delta\lambda(T)$
measured on Film~I with $T_c=30$~K shows similar behavior
demonstrating weak coupling regime at low temperatures and steep
slope in $R_s(T)$. These observations support the hypothesis of
the presence of the small gap in this film.

In contrast, Film II shows no sign of the smaller gap in the
ab-plane. Extracted value $\Delta(0)=7.4\pm 0.25$ meV is in good
agreement with reported values of the larger gap (see Table I) and
maximum gap measured in a c-axis orientated film with $T_c=38$~K
\cite{Chen}.

There are two possible explanations. The first is the presence of
an anisotropic gap $\Delta(\vec k)=\Delta_{ab} cos(\theta)
+\Delta_{c}sin(\theta)$ that comes from the 2D Fermi surface,
where $\Delta_{ab}\sim 7.5$ meV and $\Delta_c\sim 3$~meV. The
presence of anisotropic gap in MgB$_2$ is discussed widely
\cite{Junod,Haas,Chen}. In the c-axis orientated film only the
large gap $\Delta_{ab}$ determines the microwave properties
whereas on unaligned samples the smaller gap $\Delta_c$ defines
the transport behavior. Differences in the form of $\lambda(T)$
for the powder sample and Film I are probably related to the
differences properties of the surface (where the gap becomes
isotropic over the directions in the momentum space
\cite{Golubov2}), and the difference of mean free path. The second
peak at $2.8\div 3.9$ meV in the (DOS) measured in the tunnel and
point contact experiments comes from a surface reduced gap.
Different kinds of surface region were observed with an STM
\cite{Giubileo1} recently. One type of region was found which
showed the large gap $\Delta=6$ meV and another region with the
small gap $\Delta=3$ meV only. This could be associated with
different crystallographic orientations in each type of region.
The scenario of an anisotropic gap with $\Delta_{ab}>\Delta_c$
contradicts the tunnelling measurements of Chen et al.,
\cite{Chen} and the high Coulomb repulsion in ab-plane suggested
in \cite{Voelker}.

A second explanation is that some samples contain an Mg deficient
phase (due to MgO formation \cite{Berenov}) that has the smaller
gap seen in the bulk samples and in the films with $T_c<38$~K. In
order to explain the heat capacity data the volume of second phase
would need to be comparable to the primary phase. This scenario
implies that Film II is single phase only.

In conclusion, the results of the temperature dependence of the
microwave surface impedance measured on three different samples
are reported. Powder sample and Film I with $T_c=30$~K show strong
temperature dependence in both components of $Z_s$ which
associated with presence of the small gap $\Delta<1.76kT_c$. Film
II (c-axis aligned, $T_c=38$~K) shows strong coupling behavior in
the $\lambda(T)$. Extracted values of the London penetration depth
and superconducting gap are $\lambda_L(0)=110\pm 10$~nm and
$\Delta(0)=7.4\pm 0.25$~meV. Two possible explanations have been
introduced in terms of either the existence of an anisotropic
superconducting gap or the presence of second phase with lower gap
in some MgB$_2$ samples. In the light of our current results the
former scenario looks much more plausible. These experimental
results show that microwave measurements on single crystals of
MgB$_2$ for the current flowing in ab and c directions are highly
desirable.

The work is supported by the EPSRC GR/M67445, GR/R55467, NPL and
the Royal Society.\\

\begin{figure}[t]
\caption{Temperature dependence of the microwave
surface resistance. Powder (solid circles), Film I (open circles)
and Film II (solid squares). } \label{Rs}
\end{figure}

\begin{figure}[t]
\caption{Temperature dependence of the field
penetration depth of the powder sample (open circles) and Film I
(solid circles). Lines are the fit using equation (1) $n=3-T/T_c$
(solid one) and $n=4$ dashed one.} \label{Xs1}
\end{figure}

\begin{figure}[t]
\caption{Temperature dependence of the field
penetration depth of the Film II (dots). Lines are the fit using
equation (1) $n=4$ (solid one) and $n=3-T/T_c$ (dashed one).
Dash-dotted line is the fit of exponential behaviour with
$\Delta/kT_c=2.25$.} \label{Xs2}
\end{figure}

\begin{table}
\caption{Measurements of the double or anisotropic superconducting
gap in MgB$_2$ with different techniques.}\label{table1}
\begin{tabular}{cccc}
Ref. & Methods & $\Delta_L$ (meV) & $\Delta_S$ (meV) \\
\tableline
\cite{Chen} & tunnelling & 8 & 5 \\
\cite{Giubileo1} & tunnelling & 7.5 & 3.9 \\
\cite{Giubileo1} & tunnelling & 7.8 & 3.8 \\
\cite{Szabo} & point contacts & 7 & 2.8 \\
\cite{Junod} & specific heat & 6.4 & 2.1 \\
\cite{Bouquet} & specific heat & 7.2 & 2.0 \\
\cite{Tsuda} & photoemission & 5.6 & 1.7 \\
\end{tabular}
\end{table}

\begin{table}
\caption{Measurements of the temperature dependence of the field
penetration depth on MgB$_2$ with different
techniques.}\label{table2}
\begin{tabular}{rcl}
Ref. & Methods & n \\
\tableline
\cite{Chen2} & AC, M(T) & 2.8 \\
\cite{Li} & M(T) (VSM,SQUID)  & 1 \\
\cite{Panagopoulos} & AC, $\mu$SR  & 2 \\
\cite{Pronin} & Optic  & 2 \\
\end{tabular}
\end{table}

\end{multicols}


\begin{references}

\bibitem{Naga} J.~Nagamatsu, N.~Nakagawa, T.~Muranaka {\it et al.}, Nature {\bf 410}, 63 (2001).

\bibitem{Rubio} G. Rubio-Bollinger, H. Suderow, S. Vieira, cond-mat/0102242

\bibitem{Bugoslavsky} Y. Bugoslavsky, G.K. Perkins, X.Qi {\it et al.}, Nature {\bf 410}, 563
(2001)

\bibitem{Giubileo1} F. Giubileo, D. Roditchev, W. Sacks {\it et al.}, cond-mat/0105146.

\bibitem{Giubileo2} F. Giubileo, D. Roditchev, W. Sacks {\it et al.}, cond-mat/0105592.

\bibitem{Szabo} P. Szabo, P. Samuely, J. Kacmarcik {\it et al.}, cond-mat/0105598.

\bibitem{Tsuda} S. Tsuda, T. Yokoya, T. Kiss {\it et al.}, cond-mat/0104489.

\bibitem{Junod} Y. Wang, T. Plackowski, A. Junod, cond-mat/0103181; A. Junod,
Y. Wang, F. Bouquet {\it et al.}, cond-mat/0106394.

\bibitem{Bouquet} F. Bouquet, R.A. Fisher, N.E. Phillips {\it et al.}, cond-mat/0104206, to be
published in PRL.

\bibitem{Haas} Stephan Haas, Kazumi Maki, cond-mat/0104207.

\bibitem{Chen} C.-T. Chen, P. Seneor, N.-C. Yeh {\it et al.}, cond-mat/0104285.

\bibitem{Moon} S. H. Moon, J. H. Yun, H. N. Lee {\it et al.},
cond-mat/0104230.

\bibitem{Jung} J. H. Jung, K. W. Kim, H. J. Lee {\it et al.}, cond-mat/0105180.

\bibitem{Chen2} X. H. Chen, Y. Y. Xue, R. L. Meng {\it et al.}, cond-mat/0103029.

\bibitem{Li} S. L. Li, H. H. Wen, Z. W. Zhao {\it et al.}, cond-mat/0103032.

\bibitem{Panagopoulos} C. Panagopoulos, B.D. Rainford, T. Xiang {\it et al.},
cond-mat/0103060.

\bibitem{Pronin} A. V. Pronin, A. Pimenov, A. Loidl {\it et al.},
cond-mat/0104291.

\bibitem{Golubov} A.A.~Golubov, M.R.~Trunin, A.A.~Zhukov {\it et al.}, JETP Lett. {\bf 62}, 496
(1995).

\bibitem{Manzano} F. Manzano, A. Carrington, cond-mat/0106166.

\bibitem{Zhukov} A.A.~Zhukov, L.F.~Cohen, K.~Yates {\it et al.}, Supercond Sci. Technol. {\bf 14} L13
(2001) (cond-mat/0103587).

\bibitem{Hakim} N. Hakim, P.V. Parimi, C. Kusko {\it et al.}, cond-mat/0103422.

\bibitem{Berenov} 1. A. Berenov, Z. Lockman, X. Qi {\it et al.}, cond-mat/0106278

\bibitem{Zhai} H.Y. Zhai, H.M. Christen, L. Zhang {\it et al.}, cond-mat/0103618

\bibitem{Paranthaman} M. Paranthaman, C. Cantoni, H. Y. Zhai {\it et al.}, cond-mat/0103569

\bibitem{Jin} R. Jin, M. Paranthaman, H.Y. Zhai {\it et al.}, cond-mat/0104411

\bibitem{Kortus} J.~Kortus, I.I.~Mazin, K.D.~Belashchenko {\it et al.}, Phys Rev. Lett. {\bf 86}, 4656 (2001)
[cond-mat/0101446 (2001)].

\bibitem{Karen} N. McN Alford et al., J. of Supercond. {\bf 10}, 467 (1997).

\bibitem{Hao} L.~Hao, J.~Gallop, A.~Purnell {\it et al.}, J. Supercond.
{\bf 14}, 31 (2000).

\bibitem{Zhukov2} A.A.~Zhukov, A.~Purnell, L.F.~Cohen {\it et al.}, preprint.

\bibitem{Hakim} N. Hakim, P.V. Parimi, C. Kusko {\it et al.}, cond-mat/0103422.

\bibitem{Lee} S.Y. Lee, J.H. Lee, J.H. Lee {\it et al.}, cond-mat/0105327.

\bibitem{Bonn} D.A.~Bonn, P.~Dosanjh, R.~Liang {\it et al.}, Phys. Rev. Lett {\bf
68} 4545 (1992).

\bibitem{Finnemore} D.K.~Finnemore, J.E.~Ostenson, S.L.~Bud'ko {\it et al.}, Phys. Rev. Lett
{\bf 84}, 2420 (2001) [cond-mat/0102114].

\bibitem{Thompson} J.R. Thompson, M. Paranthaman, D.K. Christen {\it et al.}, cond-mat/0103514.

\bibitem{Golubov2} A.A.~Golubov and M.Yu.~Kupriyanov, JETP Lett.
{\bf 69}, 262 (2001)

\bibitem{Voelker} K.~Voelker, V.I.~Anisimov, T.M.~Rice, cond-mat/0106082.

\end{references}
\end{document}